# Magnetic ordering of the Mo$_3$O$_8$-type cluster Mott insulator Na$_3$Sc$_2$(MoO$_4$)$_2$Mo$_3$O$_8$ with spin-1/2 triangular lattice prepared via optimal synthesis

Yuya Haraguchi*, Daigo Ishikita, and Hiroko Aruga Katori
*Department of Applied Physics and Chemical Engineering, Tokyo University of Agriculture and Technology, Koganei, Tokyo 184-8588, Japan*
*Corresponding author: chiyuya@go.tuat.ac.jp

We detail the optimized synthesis of the Mo$_3$O$_8$-type cluster Mott insulator (CMI) Na$_3$Sc$_2$(MoO$_4$)$_2$Mo$_3$O$_8$, which has been considered a candidate for realizing the spin liquid ground state. The optimized Na$_3$Sc$_2$(MoO$_4$)$_2$Mo$_3$O$_8$, characterized by x-ray diffraction, energy-dispersive x-ray spectroscopy, and magnetic and heat capacity measurements, exhibited an effective magnetic moment close to the ideal 1.73 $\mu_B$ for $S$ = 1/2 spin and magnetic ordering at ~ 5 K. These observations categorize Na$_3$Sc$_2$(MoO$_4$)$_2$Mo$_3$O$_8$ as the second Mo$_3$O$_8$-type CMI to achieve a magnetic ground state, following Li$_2$InMo$_3$O$_8$. They highlight the stabilization of the magnetic ground state over the theoretically anticipated quantum spin liquid state through precise valence and chemical disorder tuning. Our findings challenge the existing theory that the magnetic ground state of Mo$_3$O$_8$-type CMIs is determined by the breathing parameter, instead showing that magnetic order is suppressed by spin defects. This study underscores the crucial role of chemical precision in investigating quantum magnetism. It suggests that precise tuning of valence states could induce magnetic ordering in previously nonmagnetic Mo$_3$O$_8$-type CMIs. Additionally, the negative findings regarding the existence of quantum spin liquids highlight the need for applied research and a reevaluation of our fundamental understanding of electronic states from both theoretical and experimental aspects.

## I.  INTRODUCTION

Quantum spin-liquids (QSLs) have garnered significant attention in condensed matter physics because of their potential to exhibit exotic quantum phenomena like fractionalized excitations and long-range entanglement [1-5]. QSL materials are notable for their highly frustrated spin configurations and the lack of conventional long-range magnetic order. These unique properties have made QSLs challenging to study, requiring innovative synthetic methods and advanced observational tools. Furthermore, discovering these materials provides new insights into quantum physics, offering valuable perspectives across various research fields [6-9].

Mo$_3$O$_8$-type cluster Mott insulators (CMIs) are promising candidates for realizing quantum spin liquids (QSLs) [10-23]. These compounds consist of Mo atoms arranged in kagome layers [Figure 1(a)], trimerized to form a triangular lattice of Mo$_3$ clusters [Figure 1(b)]. The magnetism of the Mo$_3$ cluster is determined by its valence state. As shown in Figure 1(c), [Mo$^{4+}$]$_3$ has no unpaired electrons and is nonmagnetic, whereas [Mo$^{3.67+}$]$_3$ has an unpaired electron in the $a_1$ molecular orbital, giving the Mo$_3$ cluster an $S$ = 1/2 spin. While most Mo$_3$O$_8$-type compounds are nonmagnetic with no unpaired electrons, recent developments have produced several reduced Mo$_3$O$_8$-type compounds with the latter valence state. These have attracted attention as a new frontier for CMI physics. Studies of LiZn$_2$Mo$_3$O$_8$ have revealed an intriguing phenomenon: the Curie constant at low temperatures is observed to be one-third of its value at high temperatures, implying the disappearance of two-thirds of the total spins [10]. An initial explanation for this partial spin disappearance mechanism was proposed, suggesting the formation of a spin-singlet on the triangular lattice of a honeycomb superlattice, with the remaining one-third of the spin constituting a valence-bond solid (VBS) [10].

Subsequently, Mo$_3$O$_8$-type CMIs were reinterpreted using the extended Hubbard model on the 1/6 filled breathing kagome lattice [14,16]. This model led to the development of a theoretical framework describing the electronic state of the system, postulating the plaquette charge order (PCO) state as the ground state. In the PCO state, three unpaired electrons are arranged in a hexagonal configuration, with each electron placed at a vertex and not adjacent to its neighbors. However, the effect of the PCO state on magnetic order is not yet understood. While it accounts for the reduced spin in LiZn$_2$Mo$_3$O$_8$, its role in the emergence of a quantum spin liquid state remains unclear, given the frustrated geometry. Moreover, in other Mo$_3$O$_8$-type CMIs, only Li$_2$InMo$_3$O$_8$ exhibits magnetic ordering [19], whereas Li$_2$ScMo$_3$O$_8$ [19], Li$_2$(In$_{1-x}$Sc$_x$)Mo$_3$O$_8$ [15], Na$_3$Sc$_2$(MoO$_4$)$_2$Mo$_3$O$_8$ [21,22], and Na$_3$In$_2$(MoO$_4$)$_2$Mo$_3$O [21] do not exhibit magnetic ordering, like LiZn$_2$Mo$_3$O$_8$. The connection between the PCO state and the suppression of magnetic order in Mo$_3$O$_8$-type CMIs is still under investigation. The assumption that spins on a triangular lattice naturally form a spin liquid state is questionable without considering extended interactions. Whether this suppression is due to the decoupling of superexchange interactions, as proposed in Flint and Lee's emergent honeycomb lattice model [11], or to other contributing dynamics, the origin of the spin liquid state in Mo$_3$O$_8$-type CMIs remains an open question.

On the other hand, a very recent report by Sandvik *et al.* revealed a significant deviation in the stoichiometric ratio of LiZn$_2$Mo$_3$O$_8$ [18]. By closely controlling the solid-state chemical method to bring the stoichiometric ratio closer to the ideal value, they found that the temperature dependence of the magnetic susceptibility described by the PCO model is not reproduced. Instead, the magnetic susceptibility is characterized by developing short-range order correlations typical of two-dimensional antiferromagnetic materials. This finding refutes the microscopic theoretical model previously discussed for Mo$_3$O$_8$-type CMIs. The magnetic susceptibility of LiZn$_2$Mo$_3$O$_8$, which has an almost perfect stoichiometric ratio, cannot be replicated by the PCO model. Instead, it exhibits characteristics consistent with a strongly correlated two-dimensional antiferromagnetic system. As the stoichiometric ratio improves, PCO-like behavior decreases, raising questions about the validity of traditional hypotheses.

**Table 1.** Na component, lattice parameters, unit cell volume, Weiss temperature, and effective magnetic moment in optimized Na-inserted sample, unoptimized Na-inserted sample, and precursor.

| sample | Na content | $a$ (Å) | $c$ (Å) | $V$ (Å$^3$) | $\theta_w$ (K) | $\mu_{eff}$ ($\mu_B$) |
|---|---|---|---|---|---|---|
| Optimized Na$_3$Sc$_2$(MoO$_4$)$_2$Mo$_3$O$_8$ | 2.98(3) | 5.79195(7) | 11.1121(2) | 322.541 | −184(3) | 1.788(6) |
| Unoptimized Na$_3$Sc$_2$(MoO$_4$)$_2$Mo$_3$O$_8$ | 2.52(3) | 5.78538(1) | 11.11523(4) | 321.788 | −97.6(2) | 1.46(1) |
| Precursor Na$_2$Sc$_2$(MoO$_4$)$_2$Mo$_3$O$_8$ | 2 | 5.7640(1) | 11.161(1) | 321.2708 | −31.3(9) | 0.359(7) |

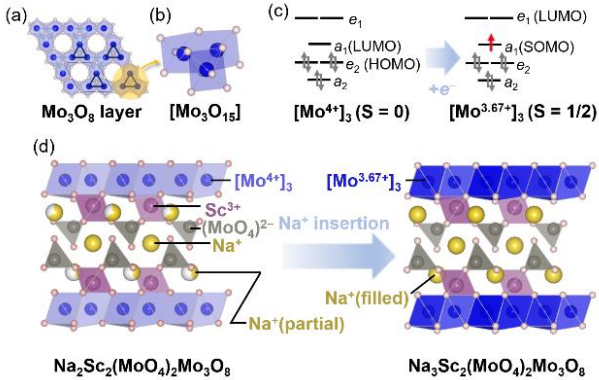

**Figure 1.** (a) Depiction of the Mo$_3$O$_8$ layer, where Mo bonds exhibiting short bond distances are illustrated in blue, and those with extended bond distances are represented in light blue. (b) The Mo$_3$O$_{15}$ unit, representing the most fundamental constituent of the Mo$_3$O$_8$ layer, encompasses Mo bonds characterized by short bond distances. (c) Elucidation of electronic coordination corresponding to molecular orbitals and their respective valence states within the [Mo$^{4+}$]$_3$ and [Mo$^{3.67+}$]$_3$ clusters. (d) Schematic representation of the synthetic process for Na$_3$Sc$_2$(MoO$_4$)$_2$Mo$_3$O$_8$ through sodium ion intercalation into the precursor Na$_2$Sc$_2$(MoO$_4$)$_2$Mo$_3$O$_8$ with their crystal structures.

Moreover, we previously reported on the detailed relationship between the structure, valence, and magnetism of Li$_2$RMo$_3$O$_8$ ($R$ = Y, Sc, Lu), synthesized by excess lithium intercalation into Li$R$Mo$_3$O$_8$ [23]. This research casts doubt on the previously accepted characteristics of Mo$_3$O$_8$-type CMIs, including the prevalence of the PCO state. Thus, we have elucidated that the valence state of the Mo$_3$ cluster is a crucial factor affecting the physical properties of Mo$_3$O$_8$-type CMIs. Considering that Li$_2$InMo$_3$O$_8$, the only Mo$_3$O$_8$-type CMI exhibiting magnetic ordering, achieves an almost ideal valence state, we propose that by appropriately adjusting the valence in other seemingly spin-deficient Mo$_3$O$_8$-type CMIs, magnetic ordering could be realized, which also clarifies that the PCO state is not an accurate physical representation of the ground state for Mo$_3$O$_8$-type CMI.

We considered Na$_3$Sc$_2$(MoO$_4$)$_2$Mo$_3$O$_8$ a suitable compound to demonstrate this hypothesis. Na$_3$Sc$_2$(MoO$_4$)$_2$Mo$_3$O$_8$ can be obtained by first synthesizing the stable nonmagnetic precursor Na$_2$Sc$_2$(MoO$_4$)$_2$Mo$_3$O$_8$, then mixing it with sodium azide NaN$_3$ and reheating in a vacuum. During this process, NaN$_3$ decomposes into Na metal and N$_2$ gas, and the Na intercalates directly into the defect sites of Na$_2$Sc$_2$(MoO$_4$)$_2$Mo$_3$O$_8$, reducing the Mo from [Mo$^{4+}$]$_3$ to [Mo$^{3.67+}$]$_3$ as shown in Fig. 1(c). Through this reduction process, as shown in Fig. 1(c) an unpaired electron is inserted into the previously empty a$_1$ orbital of the molecular orbitals formed by the Mo$_3$ cluster. As a result, each Mo$_3$ cluster has one unpaired electron, exhibiting $S = 1/2$ magnetism. However, there may be instances where the Na insertion is not fully completed during this topochemical process. Indeed, the effective magnetic moment in previous samples was estimated to be approximately 1.4 $\mu_B$ [19]. Considering that an ideal $S = 1/2$ spin should have a magnetic moment of 1.73 $\mu_B$, this suggests a deviation from the perfect Mo valence state of [Mo$^{3.67+}$]$_3$. This deviation indicates that not all Mo$_3$ cluster orbitals contain an $S = 1/2$ unpaired electron spin, as illustrated in Fig. 1(d). Additionally, Na$_3$Sc$_2$(MoO$_4$)$_2$Mo$_3$O$_8$ has also been synthesized by the solid-state reaction method, but the effective magnetic moment is even smaller at ~1.2 $\mu_B$ [22], suggesting a further deviation of the valence state of [Mo$^{3.67+}$]$_3$.

In this study, we aimed to optimize the Na intercalation process to achieve complete Na insertion into Na$_2$Sc$_2$(MoO$_4$)$_2$Mo$_3$O$_8$, introduce $S = 1/2$ unpaired electron spins into all Mo$_3$ cluster orbitals, and elucidate the magnetic properties of the sample to reveal the true ground state of Mo$_3$O$_8$-type CMIs. By adding an excess amount of NaN$_3$ during the reductive intercalation process, ensuring that no decomposition reaction occurs beyond the stoichiometric ratio, we confirmed a change in the lattice constant indicating further Na insertion through XRD. Rietveld analysis and EDX measurements revealed that Na almost completely compensated for the missing sites. Furthermore, magnetic susceptibility and specific heat measurements of the obtained samples indicated that all Mo$_3$ clusters show an effective magnetic moment of approximately 1.73$\mu_B$, suggesting successful reduction to have unpaired electrons, and exhibited magnetic ordering at ~ 5 K. Our study thus elucidates the incompleteness of the spin-lattice model from a different perspective, particularly in frustrated materials such as CMIs, and demonstrates that this can lead to erroneous conclusions about the weak magnetic ground state of this exemplary geometrically frustrated model.

**Table 2.** Crystallographic parameters for an optimized sample of $Na_3Sc_2(MoO_4)_2Mo_3O_8$ with a space group of $P\bar{3}m1$ determined from powder x-ray diffraction experiments. The obtained lattice parameters are $a$ = 5.79196(2) Å and $c$ = 11.10213(7) Å. $B$ is the atomic displacement parameter.

| atom | site | $x$ | $y$ | $z$ | $B$ (Å) |
|---|---|---|---|---|---|
| Sc1 | 1b | 1/3 | 2/3 | 0.52649(2) | 0.50(8) |
| Sc2 | 1b | 1/3 | 2/3 | 0.93299(2) | 0.12(7) |
| Mo1 | 1a | 0 | 0 | 0.3611(1) | 1.10(4) |
| Mo2 | 1c | 0 | 0 | 0.1067(1) | 1.22(4) |
| Mo3 | 3d | 0.85085(9) | 1−$x$ | 0.72533(2) | 1.48(1) |
| Na1 | 1b | 1/3 | 2/3 | 0.2448(8) | 2.4(1) |
| Na2 | 1c | 2/3 | 1/3 | 0.4755(5) | 0.9(1) |
| Na3 | 1a | 0 | 0 | 0.02146(4) | 1.2(1) |
| O1 | 3d | 0.1709(5) | 1−$x$ | 0.8427(3) | 1.6(2) |
| O2 | 3d | 0.4934(3) | 1−$x$ | 0.0596(2) | 0.6(1) |
| O3 | 1a | 0 | 0 | 0.18133(8) | 1.8(3) |
| O4 | 1c | 2/3 | 1/3 | 0.8125(6) | 1.4(3) |
| O5 | 3d | 0.5038(5) | 1−$x$ | 0.6137(3) | 1.2(1) |
| O6 | 3d | 0.1824(5) | 1−$x$ | 0.4080(3) | 1.1(1) |
| O7 | 1c | 2/3 | 1/3 | 0.2652(8) | 1.3(4) |
| O8 | 1a | 0 | 0 | 0.6110(6) | 1.3(3) |

## II. EXPRIMENTAL PROCEDURE

As delineated in prior research, the precursor $Na_2Sc_2(MoO_4)_2Mo_3O_8$ was synthesized utilizing a solid-state reaction technique. The corresponding chemical reaction is represented as follows:

$$1.2Na_2MoO_4 + Sc_2O_3 + Mo + 2.8MoO_3 \rightarrow Na_2Sc_2(MoO_4)_2Mo_3O_8 + 0.2Na_2O. \quad (1)$$

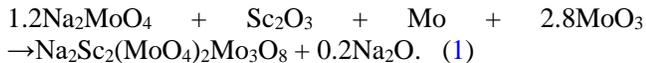

Accounting for the potential evaporation of $Na_2MoO_4$ during calcination, the $Na_2O$ quantity was adjusted to ensure precipitation as a byproduct. The materials were weighed and mixed according to the reaction (1), pelletized, encased in copper foil, and vacuum-sealed within quartz tubes. Calcination was executed at 600 °C for 60 hours. Post-calcination, the pellets were rinsed using deionized water to eliminate the $Na_2O$ byproduct. $Na_3Sc_2(MoO_4)_2Mo_3O_8$ was synthesized by intercalating sodium

into the precursor. The anticipated chemical reaction is outlined below:

$$Na_2Sc_2(MoO_4)_2Mo_3O_8 + NaN_3 \rightarrow Na_3Sc_2(MoO_4)_2Mo_3O_8 + 1.5N_2 \quad (2)$$

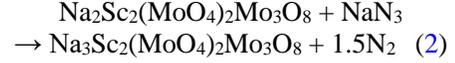

Based on the stoichiometric ratio following the reaction (2), the current research employed four times the quantity of $NaN_3$ relative to the precursor to ensure adequate sodium intercalation.

The mixture was pelletized, sealed in Pyrex tubes under an argon atmosphere, and reacted at 400 °C for 80 hours within a circular electric furnace. To avert Pyrex tube explosion due to significant nitrogen gas generation from the rapid decomposition of $NaN_3$, the temperature was gradually increased at a rate of 5 °C/h from 180 °C to 400 °C. Following the reaction, the pellets were rinsed with methanol to remove any residual elemental sodium produced by the decomposition of $NaN_3$. Moreover, it was found that adding $NaN_3$ in further excess causes the reduction reaction to proceed significantly, leading to the precipitation of Mo elements.

The obtained polycrystalline samples were characterized by powder x-ray diffraction (XRD) experiments in a diffractometer with Cu-K$\alpha$ radiation, and chemical analysis was conducted using a scanning electron microscope (JEOL IT100) equipped with an energy dispersive x-ray spectroscope (EDX with 15 kV, 0.8 nA, 1-μm beam diameter). The cell parameters and crystal structure were refined by the Rietveld method using the Z-Rietveld software [24,25]. The temperature dependence of the magnetization was measured under several magnetic fields using the magnetic property measurement system (MPMS; Quantum Design) at Institute for Solid State Physics (ISSP), the University of Tokyo. The temperature dependence of the heat capacity was measured using the conventional relaxation method in a physical property measurement system (PPMS; Quantum Design) at ISSP.

## III. RESULTS

Figure 2 shows powder XRD patterns of $Na_3Sc_2(MoO_4)_2Mo_3O_8$ after rinsing with methanol. Except for impurity origin, all peaks can be indexed by the space group $P\bar{3}m1$, characterized by trigonal lattice constants $a$ = 5.791 95(7) Å and $c$ = 11.1121(2) Å. Table 1 presents a comparison of the lattice constants for the precursor, the intercalated sample under stoichiometric $NaN_3$ conditions, and the intercalated samples under excess $NaN_3$ conditions. The sample with excess $NaN_3$ shows a larger change in lattice constant from the precursor than the stoichiometric $NaN_3$ sample, suggesting a greater amount of Na insertion. These results indicate that $NaN_3$ was mixed in excess and subsequently heat-treated to control the level of Na insertion more precisely. From this point forward, stoichiometric $NaN_3$ samples and $NaN_3$ excess samples will be referred to as "unoptimized $Na_3Sc_2(MoO_4)_2Mo_3O_8$" and "optimized

Na$_3$Sc$_2$(MoO$_4$)$_2$Mo$_3$O$_8$", respectively. The EDX measurement results show that the amount of Na in optimized Na$_3$Sc$_2$(MoO$_4$)$_2$Mo$_3$O$_8$ is approximately 2.98(3), confirming that Na is fully compensated within the standard deviation range. In contrast, the Na content in unoptimized Na$_3$Sc$_2$(MoO$_4$)$_2$Mo$_3$O$_8$ is 2.52(3), indicating that Na is only about half compensated, which suggests that the intercalation was insufficient. The optimized crystal structure of Na$_3$Sc$_2$(MoO$_4$)$_2$Mo$_3$O$_8$ underwent refinement using the Rietveld method, as detailed in the Experimental Methods section. The refined crystallographic parameters are summarized in Table 2. Rietveld analysis also confirmed that the occupation at each Na site converged to 1 within a standard deviation.

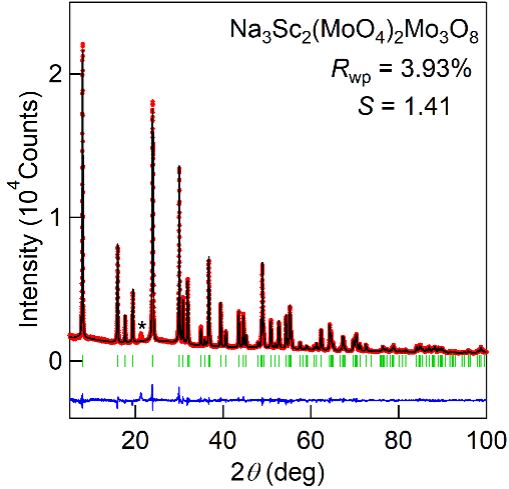

**Figure 2.** Rietveld analysis results on optimized Na$_3$Sc$_2$(MoO$_4$)$_2$Mo$_3$O$_8$ sample. Observed intensities (red), calculated intensities (black), and their differences (blue) from the Rietveld refinement. Vertical bars indicate the positions of Bragg reflections. The asterisk indicates an unknown impurity.

Figure 3(a) presents the temperature dependence of the reciprocal magnetic susceptibility $1/\chi$ for optimized Na$_3$Sc$_2$(MoO$_4$)$_2$Mo$_3$O$_8$, measured at a magnetic field of $\mu_0 H = 7$ T. For comparison, the $1/\chi$-data of unoptimized Na$_3$Sc$_2$(MoO$_4$)$_2$Mo$_3$O$_8$ is also presented. The Curie-Weiss fitting results for the respective $\chi$ data in the 150-300 K range are shown in Table 1. The unoptimized Na$_3$Sc$_2$(MoO$_4$)$_2$Mo$_3$O$_8$ sample has an effective magnetic moment of 1.31 $\mu_B$, considerably lower than the expected value of 1.73 $\mu_B$ for an $S = 1/2$ spin. This reduction suggests that the valence of the Mo$_3$ cluster is shifted from [Mo$^{3.67+}$]$_3$ due to insufficient Na ion insertion. Furthermore, the value of 1.31 $\mu_B$ indicates that only about 57% of the $S = 1/2$ spins are present, which approximately aligns with the EDX results. In contrast, the effective magnetic moment of the optimized Na$_3$Sc$_2$(MoO$_4$)$_2$Mo$_3$O$_8$ sample is estimated to be 1.788 $\mu_B$, indicating that the charge state of [Mo$_3$]$^{11+}$ with $S = 1/2$ is nearly achieved through the optimization of Na ion insertion. These results are in good agreement with XRD and EDX results. In addition, the absolute $\theta_W$-value increases significantly for the optimized sample compared to the unoptimized sample. This increase is likely due to the higher concentration of $S = 1/2$ spin on each Mo$_3$ cluster, which results in more superexchange interaction paths within the optimized sample.

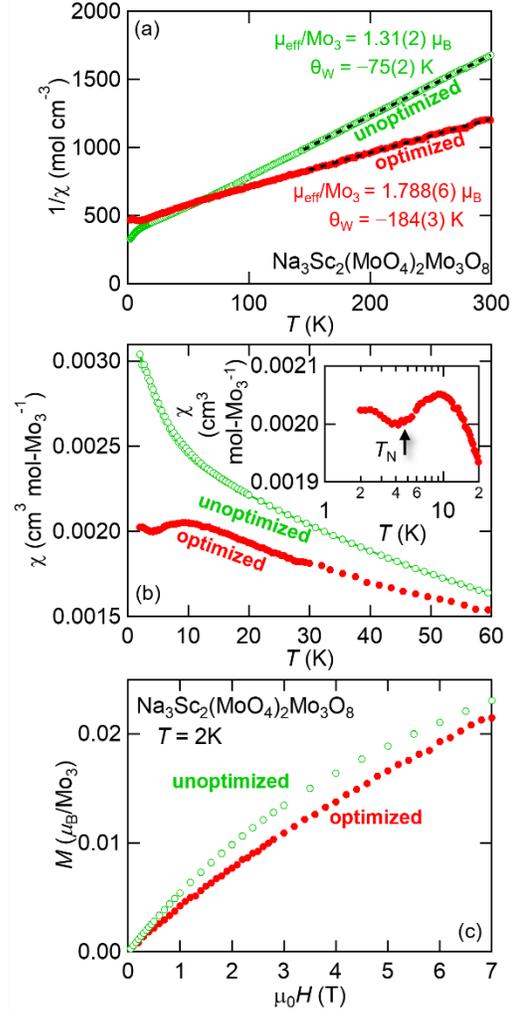

**Figure 3** (a) Temperature dependence of inversed magnetic susceptibility $1/\chi$ of optimized and unoptimized samples Na$_3$Sc$_2$(MoO$_4$)$_2$Mo$_3$O$_8$. The dashed lines on the data represent the results that fit the Curie-Weiss (CW) models. (b) Temperature dependence of $\chi$ data of optimized and unoptimized samples Na$_3$Sc$_2$(MoO$_4$)$_2$Mo$_3$O$_8$ below 60 K. The inset shows the semilogarithmic plot. (c) Isothermal magnetization $M$ at $T = 2$ K. The dashed lines on the data represent the fitting results to the Brillouin function.

Figure 3(b) depicts the temperature-dependent magnetic susceptibility $\chi$ in the low-temperature regime. For the unoptimized Na$_3$Sc$_2$(MoO$_4$)$_2$Mo$_3$O$_8$, a noticeable divergence in $\chi$ begins at approximately 10 K. This divergence is ascribed to partial spin loss, a phenomenon that has been extensively reported in various Mo$_3$O$_8$-type magnets. The underlying mechanism is believed to involve two distinct Curie-Weiss behaviors stemming from the partial spin defect. In contrast, optimized $Na_3Sc_2(MoO_4)_2Mo_3O_8$ demonstrates a magnetic anomaly at ~5 K, likely indicative of magnetic ordering. Such a temperature dependence of χ is also observed in $Li_2InMo_3O_8$, the sole $Mo_3O_8$-type magnet confirmed to exhibit magnetic ordering, highlighting a unique aspect of its magnetic behavior.

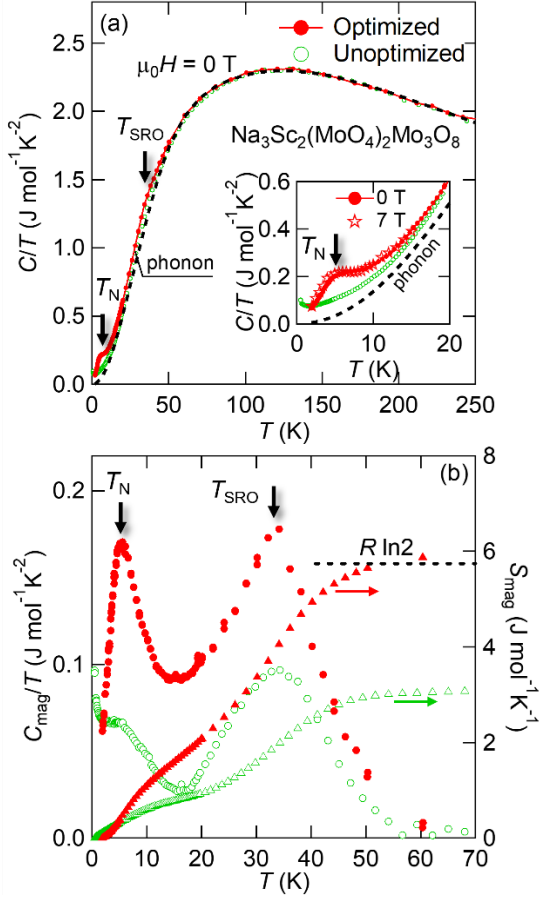

**Figure 4.** (a) Temperature dependence of the heat capacity divided by temperature $C/T$ in optimized and unoptimized $Na_3Sc_2(MoO_4)_2Mo_3O_8$ samples, compared with the nonmagnetic precursor $Na_2Sc_2(MoO_4)_2Mo_3O_8$ to estimate the phonon contribution ($C_{phonon}/T$). The inset highlights the low-temperature region. (b) The magnetic heat capacity ($C_{mag}$) after subtracting $C_{phonon}$ and the magnetic entropy ($S_{mag}$) calculated by integrating $C_{mag}/T$ with respect to $T$.

Figure 3(c) displays the isothermal magnetization process at 2 K. An upward convex magnetization process is observed in unoptimized $Na_3Sc_2(MoO_4)_2Mo_3O_8$, clearly attributed to free spins. Conversely, the magnetization process in optimized $Na_3Sc_2(MoO_4)_2Mo_3O_8$ is nearly linear. The observed differences in the magnetization processes can be attributed to the higher concentration of free spins in the unoptimized $Na_3Sc_2(MoO_4)_2Mo_3O_8$ compared to the optimized $Na_3Sc_2(MoO_4)_2Mo_3O_8$. This observation aligns with the enhanced Curie-tail near 0 K in the χ-data of the unoptimized $Na_3Sc_2(MoO_4)_2Mo_3O_8$, as shown in Fig. 3(b). Determining whether the free spin component results from emergent isolated spins associated with forming a valence bond glass or simply from impurities is challenging in the present stage. Nonetheless, the effective magnetic moment appears ideal, and the free-spin component is significantly suppressed.

Figure 4(a) showcases the heat capacity divided by temperature $C/T$, with an inset providing a detailed view of the low-temperature segment. In the optimized $Na_3Sc_2(MoO_4)_2Mo_3O_8$, the $C/T$ plot reveals a pronounced λ-shaped peak at the Néel temperature $T_N$ of approximately 5 K. This peak corresponds to the temperature at which magnetic anomalies were previously identified in the χ-data presented in Fig. 3(b). Moreover, as shown in the inset of Fig. 4(a), applying a magnetic field causes the $C/T$ peak to shift towards a lower temperature, a hallmark of the antiferromagnetic transition. In contrast, the $C/T$ data for unoptimized $Na_3Sc_2(MoO_4)_2Mo_3O_8$ exhibit no significant features. These observations suggest that the optimized $Na_3Sc_2(MoO_4)_2Mo_3O_8$ undergoes a second-order magnetic phase transition that is bulk at $T_N$. Additionally, broad, hump-like structures observed in the $C/T$ data around 35 K likely indicate short-range magnetic order.

Surprisingly, the specific heat at temperatures above 100 K is nearly identical for optimized $Na_3Sc_2(MoO_4)_2Mo_3O_8$, unoptimized $Na_3Sc_2(MoO_4)_2Mo_3O_8$, and precursor $Na_2Sc_2(MoO_4)_2Mo_3O_8$. This similarity indicates that the phonons in these samples remain almost constant and are largely unaffected by the minor presence of Na defects. Here, by considering the total heat capacity of $Na_2Sc_2(MoO_4)_2Mo_3O_8$ as the phonon contribution of heat capacity in $Na_3Sc_2(MoO_4)_2Mo_3O_8$ and subtracting it, we extracted the magnetic specific heat of $Na_3Sc_2(MoO_4)_2Mo_3O_8$. Figure 4(b) displays the magnetic contribution to the heat capacity divided by temperature $C_{mag}/T$, derived by subtracting the lattice contribution from the experimental data. Additionally, the magnetic entropy $S_{mag}$ is calculated by integrating $C_{mag}/T$ with respect to $T$, assuming $C_{mag}/T$ equals 0 at 0 K following the third law of thermodynamics. For optimized $Na_3Sc_2(MoO_4)_2Mo_3O_8$, the high-temperature asymptotic value of $S_{mag}$ approaches $R\ln 2$ = 5.76 J/mol-K, indicating that nearly all $Mo_3$ clusters exhibit an $S = 1/2$ spin, resulting from the realization of an ideal $[Mo^{3.67+}]_3$ valence state. In contrast, the $S_{mag}$ values for unoptimized $Na_3Sc_2(MoO_4)_2Mo_3O_8$ are significantly lower, suggesting a deviation from the $[Mo^{3.67+}]_3$ valence. This deviation is likely due to the loss of Na ions, corresponding to a decrease in the effective magnetic moment.

The temperature dependence of $C_{mag}/T$ reveals that the contribution from the magnetic specific heat, associated with short-range order around $T_{SRO}$ ~ 35 K, becomes increasingly significant. Interestingly, a comparable feature near $T_{SRO}$ ~ 35 K is also observed in the $C_{mag}/T$ profile of unoptimized $Na_3Sc_2(MoO_4)_2Mo_3O_8$. This observation suggests the presence of short-range order influenced by the low-dimensional nature of spin correlations; an attribute that appears to be independent of the concentration of spins. This phenomenon underscores the intrinsic properties of the system that may drive the observed magnetic behavior, highlighting the critical role of spin correlations in determining the magnetic specific heat near $T_\text{SRO}$.

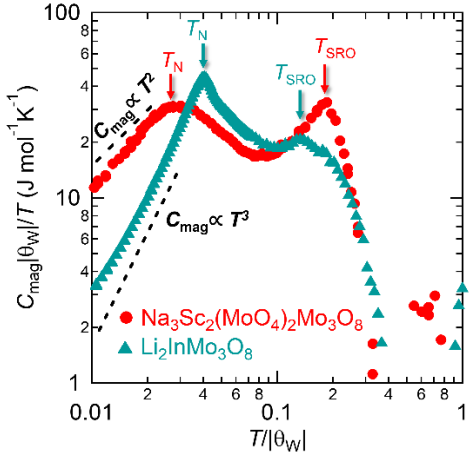

**Figure 5.** The normalized magnetic specific heat of $Na_3Sc_2(MoO_4)_2Mo_3O_8$, compared with the frustrated $Mo_3O_8$-type CMI cluster magnet $Li_2InMo_3O_8$, shows magnetic ordering, with $T$ normalized by the absolute Curie-Weiss temperature $|\theta_W|$ for comparison.

## IV. DISCUSSION

As we previously reported, $Na_3Sc_2(MoO_4)_2Mo_3O_8$ was synthesized by insertion of Na ions at Na-deficient sites of nonmagnetic $Na_2Sc_2(MoO_4)_2Mo_3O_8$ using a decomposition reaction of $NaN_3$ [21]. By introducing $NaN_3$ in large excess stoichiometry, ensuring that the sample does not decompose, it was possible to complete Na-insertion. The resulting sample exhibited an effective magnetic moment, supporting the conclusion that the $Mo_3$ cluster displays a spin of 1/2. This result suggests that the replenishment of Na in previous stoichiometric mixtures might have been impure, as supported by EDX measurements. Additionally, susceptibility and specific heat measurements clarified that $Na_3Sc_2(MoO_4)_2Mo_3O_8$ exhibits magnetic ordering at approximately 5 K. This makes it the second material, following $Li_2InMo_3O_8$ [19], to achieve a magnetically ordered state among $Mo_3O_8$-based cluster Mott insulators.

To elucidate the similarities and differences in the magnetic properties of $Li_2InMo_3O_8$ and $Na_3Sc_2(MoO_4)_2Mo_3O_8$, data on magnetic heat capacity normalized by the Weiss temperature is shown in Fig. 5. Both compounds exhibit a double peak corresponding to long-range order ($T_N$) and short-range order ($T_\text{SRO}$), which appear similar at first glance but reveal differences upon closer inspection. The magnetic ordering in $Na_3Sc_2(MoO_4)_2Mo_3O_8$ occurs in a lower energy region than in $Li_2InMo_3O_8$. In contrast, short-range correlations are developed in a higher energy region in $Na_3Sc_2(MoO_4)_2Mo_3O$ than in $Li_2InMo_3O_8$. The magnetic correlations between the layers of both compounds can explain this difference. The magnetic layer distance in $Li_2InMo_3O_8$ is 5.144 Å [21], whereas in $Na_3Sc_2(MoO_4)_2Mo_3O$, it is more than twice as large at 11.10 Å. Thus, it is believed that the difference in magnetic dimensionality arises between the two compounds.

Further evidence of the difference in magnetic dimensionality can be seen in the temperature dependence of $C_\text{mag}$ below $T_N$. In antiferromagnetic materials, the $C_\text{mag}$ data generally follows the equation $C_\text{mag} = A_\text{mag}T^d$ below $T_N$, corresponding to the dimensionality of magnon excitations [26], where $A_\text{mag}$ is a constant, and $d$ (= 1, 2, or 3) is the dimensionality. As shown in Fig. 5, $C_\text{mag}/T$ in $Na_3Sc_2(MoO_4)_2Mo_3O_8$ is almost linear with temperature, suggesting that $C_\text{mag}$ is directly proportional to $T^2$, corresponding to two-dimensional magnon excitations. In contrast, $C_\text{mag}/T$ in $Li_2InMo_3O_8$ is not linear but proportional to $T^2$, indicating that $C_\text{mag}$ is directly proportional to $T^3$, corresponding to three-dimensional magnon excitations. These findings demonstrate that the magnetic dimensionality differs significantly between the two materials due to the difference in magnetic layer distances, suggesting that a low dimensionality is well realized in $Na_3Sc_2(MoO_4)_2Mo_3O_8$.

Finally, we discuss the validity of the theory that the magnetic ground state of $Mo_3O_8$-type CMIs is predominantly determined by the breathing parameter $\lambda_B$, defined as the ratio of the intercluster Mo-Mo distance ($d_\text{inter}$) to the intracluster Mo-Mo bond distance ($d_\text{intra}$), with $\lambda_B = d_\text{inter}/d_\text{intra}$. Our investigation, which includes the newly synthesized optimized $Na_3Sc_2(MoO_4)_2Mo_3O_8$ with an effective magnetic moment close to the ideal 1.73 $\mu_B$ for $S = 1/2$ spins, reveals magnetic order. This finding has prompted reevaluating and modifying the relationship between the breathing parameter and the effective magnetic moment in $Mo_3O_8$-type CMIs, as illustrated in Fig. 6. The optimized $Na_3Sc_2(MoO_4)_2Mo_3O_8$ has a $\lambda_B$ value of approximately 1.22, nearly identical to that of the unoptimized $Na_3Sc_2(MoO_4)_2Mo_3O_8$. Additionally, Fig. 6 demonstrates a minimal correlation between the $\lambda$ value and the presence of magnetic order. Conversely, optimized $Na_3Sc_2(MoO_4)_2Mo_3O_8$ and $Li_2InMo_3O_8$ exhibit magnetic order and have effective magnetic moments close to 1.73 $\mu_B$. In contrast, most $Mo_3O_8$-type compounds that do not exhibit magnetic order have significantly smaller effective magnetic moments.

These findings suggest that the magnetic state is not governed by $\lambda_B$. Instead, we conclude that magnetic order is suppressed by spin defects, which can be parameterized by the deviation of the effective magnetic moment from 1.73 $\mu_B$. This result strongly indicates that the previously proposed theory for $Mo_3O_8$-type CMIs needs to be reconsidered [14,16]. In other words, the lack of magnetic ordering in previously reported $Mo_3O_8$-type CMIs can be attributed to a spin defect caused by a valence shift from ideal $[Mo^{3.67+}]_3$. Very recently, a theoretical model proposed by Watanabe et al. suggests that quenched randomness in the exchange interaction of $S = 1/2$ triangular-lattice Heisenberg antiferromagnets leads to a QSL-like state [28], explaining experimental observations in magnetically disordered $Mo_3O_8$-type CMIs. We propose that $Mo_3O_8$-type CMIs, which exhibit small effective magnetic moments and lack magnetic ordering, can be interpreted as exhibiting magnetic disorder within the framework of this model.

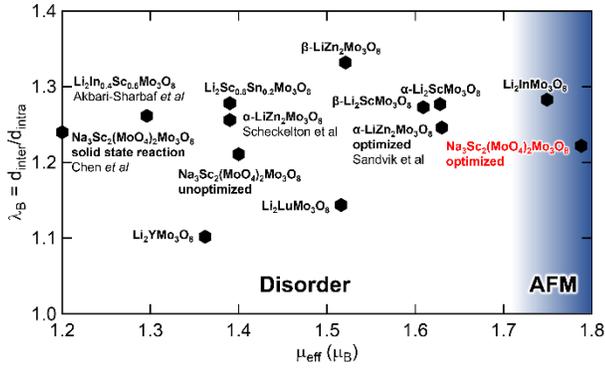

**Figure 6.** Relation between effective magnetic moment, breathing parameter λ, and magnetic states in $Mo_3O_8$-type CMIs. The other plotted data are reproduced from Refs. 13, 15, 18, 19, 21-23, 27.

Furthermore, it suggests that even for $Mo_3O_8$-type CMIs that have not previously exhibited magnetic order, there is potential to induce magnetic order through precise adjustment of the valence state. For example, in $LiZn_2Mo_3O_8$, Sandvik et al. found significant sample dependency due to composition variation [18]. Although the sample closest to the ideal composition does not exhibit magnetic order, it does show strong short-range interlayer interactions characteristic of a two-dimensional antiferromagnet. Therefore, optimization of synthesis methods is also highly desirable for other $Mo_3O_8$-type CMIs.

## V. SUMMARY

We successfully optimized the synthesis of $Na_3Sc_2(MoO_4)_2Mo_3O_8$, yielding a sample with an effective magnetic moment close to the ideal 1.73 $\mu_B$ for $S$ = 1/2 spins. Magnetic susceptibility and specific heat measurements confirmed magnetic ordering at approximately 5 K, making $Na_3Sc_2(MoO_4)_2Mo_3O_8$ the second $Mo_3O_8$-type CMI to achieve this state, following $Li_2InMo_3O_8$. Our findings challenge the theory that the magnetic ground state of $Mo_3O_8$-type CMIs is determined by the breathing parameter, showing instead that magnetic order is suppressed by spin defects, indicated by deviations from the ideal effective magnetic moment of 1.73 $\mu_B$. This study suggests revising the existing theory for $Mo_3O_8$-type CMIs and indicates that precise valence state adjustment can potentially induce magnetic order in previously reported $Mo_3O_8$-type CMIs without magnetic orderings.

## ACKNOWLEDGMENT

This work was supported by JST PRESTO Grant Number JPMJPR23Q8 (Creation of Future Materials by Expanding Materials Exploration Space) and JSPS KAKENHI Grant Numbers. JP23H04616 (Transformative Research Areas (A) "Supra-ceramics"), JP24H01613 (Transformative Research Areas (A) "1000-Tesla Chemical Catastrophe"), JP22K14002 (Young Scientific Research), and JP24K06953 (Scientific Research (C)). Part of this work was carried out by joint research in the Institute for Solid State Physics, the University of Tokyo (Project Numbers 202012-HMBXX-0021, 202112-HMBXX-0023, 202106-MCBXG-0065 and 202205-MCBXG-0063).